\documentclass[prl,twocolumn,showpacs,epsfig,preprintnumbers,amsmath,amssymb]{revtex4}
\usepackage{graphicx}% Include figure files
\usepackage{dcolumn}% Align table columns on decimal point
\usepackage{bm}% bold math
\def \half{ \frac {1}{2} }
\def \be{\begin{equation}}
\def \ee{\end{equation}}
\def \bea{\begin{eqnarray}}
\def \eea{\end{eqnarray}}
\def \ba{\begin{array}}
\def \ea{\end{array}}
\def \non{\nonumber}
\begin{document}

%%%%%%%%%%%%%%%%%%%%%%%%%%%%%%%%%%%%%%%%%%%%%%%%%%%%%%%%%%%%%%%%%%%%%%%%%
%%%%%%%%%%%%%%%%%%%%%%%%%%%%%%%%%%%%%%%%%%%%%%%%%%%%%%%%%%%%%%%%%%%%%%%%%

\title{An exactly solvable model for driven dissipative systems}
\author{Yair Srebro and Dov Levine}
\affiliation{Department of Physics, Technion, Haifa 32000, Israel}
\date{\today}

\begin{abstract}

We introduce a solvable stochastic model inspired by granular
gases for driven dissipative systems. We characterize far from
equilibrium steady states of such systems through the
non-Boltzmann energy distribution and compare different measures
of effective temperatures. As an example we demonstrate that
fluctuation-dissipation relations hold, however with an effective
temperature differing from the effective temperature defined from
the average energy.

\end{abstract}

%05.70.Ln Nonequilibrium and irreversible thermodynamics
%02.50.Ey Stochastic processes
%45.70.-n Granular systems

 %Use showkeys class option if keyword display desired
\maketitle

%%%%%%%%%%%%%%%%%%%%%%%%%%%%%%%%%%%%%%%%%%%%%%%%%%%%%%%%%%%%%%%%%%%%%%%%%
%Introduction:
%%%%%%%%%%%%%%%%%%%%%%%%%%%%%%%%%%%%%%%%%%%%%%%%%%%%%%%%%%%%%%%%%%%%%%%%%

Dissipative many-particle systems are far from thermodynamic
equilibrium, and a general theoretical description of their
statistical mechanics is lacking, in contrast to systems in
equilibrium, for which there is a well established theory. In this
Letter we propose a simple exactly solvable model, which provides
a context in which certain questions concerning driven dissipative
systems may be resolved unambiguously, such as whether different
proposed ``definitions'' of temperature give the same value, as is
the case for thermal equilibrium.

%Previous research on this subject has been numerical or
%approximate, and may thus give apparent agreement due to numerical
%error.

Granular materials have in recent years been considered a paradigm
for dissipative open systems \cite{granular_gases}. In such
systems comprised of macroscopic particles, energy is dissipated
via interactions, being transferred from macroscopic degrees of
freedom (motion of grains) into microscopic degrees of freedom,
and can not be transformed back. Continuous driving is needed in
order to maintain such a system in a dynamic state. This driving
may be realized, for instance, by gravity-driven flow down an
incline \cite{incline_flow}, by continuous avalanches in a
rotating drum \cite{rotating_drum}, or in a vibrated container
\cite{vibrated_container}.

An oft-studied system is a homogeneously heated granular gas
\cite{noije_ernst_1998}: a collection of hard spheres (or disks in
$2D$) undergoing inelastic collisions and driven by a stochastic
thermostat. This yields Langevin dynamics including a random
uncorrelated force $\overrightarrow{F}(t)$, satisfying $\langle
\overrightarrow{F} \rangle = 0$ and $\langle F_i(t) F_j(t')
\rangle = 2 \gamma T_B \delta (t-t') \delta_{ij}$, and a drag
force $-\gamma \overrightarrow{v}$. This may be interpreted as
coupling to a heat bath of temperature $T_B$ with a coupling
strength $\gamma$. Were the system non-dissipative, it would reach
equilibrium with temperature $T_B$, irrespective of the details of
the coupling, that is, independent of $\gamma$.

A driven dissipative system reaches a non-equilibrium steady
state, for which a granular temperature $T_G$ may be defined as
the average kinetic energy of the grains. Not only is $T_G$ always
smaller than $T_B$, but its value depends on the details of the
coupling with the bath \cite{geisshirt_1998,cecconi_2004}. This
steady state behaves statistically differently than an equilibrium
state at the same effective temperature. The energy distribution
deviates from the Boltzmann distribution, exhibiting overpopulated
high energy tails \cite{granular_gases}. Surprisingly, even though
driven dissipative systems are not in equilibrium,
fluctuation-dissipation (FD) relations often hold and may serve to
define an effective temperature, $T_{FD}$ \cite{t_fd_other}, which
has been found in numerical experiments to coincide with $T_G$
\cite{t_fd_gran}.

Granular gases may be described theoretically by the Boltzmann
equation of kinetic theory, however solutions exist only in terms
of approximations valid for small deviations from the equilibrium
Boltzmann distribution, that is for small inelasticity and low
volume fraction \cite{granular_gases}. A dissipative model for
which some exact results have been found is the one-dimensional
Maxwell-model \cite{maxwell_model}, in which particles collide
inelastically with a uniform collision rate, independent of
velocity or location. Energy distributions have also been
investigated numerically in a two-dimensional version of this
model assuming a random impact parameter in every collision
\cite{mackintosh_2004}.

In this Letter we present a novel exactly solvable dissipative
model, in which interactions occur randomly and redistribute
energy randomly between the interacting particles. Our main
results are:

$\bullet$ All moments of the model may be computed exactly.

$\bullet$ In the maximally dissipative limit, the generating
function of the model may be solved exactly.

$\bullet$ FD relations hold with $T_{FD} > T_G$.

%%%%%%%%%%%%%%%%%%%%%%%%%%%%%%%%%%%%%%%%%%%%%%%%%%%%%%%%%%%%%%%%%%%%%%%%%
%System definition:
%%%%%%%%%%%%%%%%%%%%%%%%%%%%%%%%%%%%%%%%%%%%%%%%%%%%%%%%%%%%%%%%%%%%%%%%%

Our model consists of a collection of $N$ particles having
energies ${E_i}$, with a constant interaction rate between any two
particles in the system. In every interaction two particles from
the system are chosen at random and their energies are summed. In
the case of conservative dynamics (analogous to elastic
collisions) this total energy is repartitioned randomly with a
uniform distribution between the two interacting particles
\cite{ulam_1980}, while for dissipative dynamics (as for inelastic
collisions with a constant restitution coefficient) only a
fraction $\alpha$ of the total energy is repartitioned between the
particles and the rest is dissipated out of the system.
Additionally, we couple the system to a heat bath so that it
reaches a steady state.

The simplicity of this model results from the fact that every
particle in it is described only by its energy, as opposed, for
example, to the $2d$ degrees of freedom per particle in a
$d$-dimensional frictionless hard sphere gas. By eliminating the
momentum and spatial variables and using solely the energy we turn
the vectorial collisions between particles into scalar
interactions, and preclude spatial correlations.

For conservative dynamics ($\alpha=1$) the system reaches an
equilibrium state with the exponential Boltzmann distribution for
the particles' energies, $p(E)=T^{-1} exp(-E/T)$, where the
temperature (measured in units of energy) equals the average
energy in the system $T = \langle E \rangle$. Dissipative dynamics
($\alpha<1$) cause energy to decay, and in order to maintain the
system in a steady state, we keep it in contact with a heat bath,
which is constructed as an infinitely large system of particles
obeying the conservative dynamics described above, kept in
equilibrium at a temperature $T_{B}$. The coupling of the
dissipative system to the bath is through conservative
interactions between a particle chosen at random from the system
and a particle chosen at random from the bath, and is
characterized by a coupling strength, $f$, which is defined as the
fraction of the particle's interaction that are with the bath out
of all its interactions (with the bath and with other particles in
the dissipative system). We will see that the steady state of the
dissipative system depends on the bath through both $T_B$ and $f$
\footnote{See also \cite{mackintosh_2004} who emphasized the
importance of other coupling details}.

The stochastic equation of motion for the evolution of the energy
of particle $i$ during an infinitesimal time step $dt$ is hence
given by
\bea E_i(t+dt)= \left\{
\begin{array}{cc}
  \underline{value}: & \underline{probability:} \\
  E_i(t) & 1- \Gamma dt \\
  z \alpha (E_i(t)+E_j(t))  & (1-f) \Gamma dt \\
  z (E_i(t)+E_{B})  & f \Gamma dt
\end{array} \right.\label{eq:dyn_rule},
\eea
where: $\Gamma$ is the interaction rate per particle per unit time
(which does not affect the steady state but only the rate of
approach to it); $j \in \{ 1 , ..., N \}$ is the index of the
particle with which particle $i$ may interact, chosen randomly at
every interaction; $z \in [0,1]$ is the fraction of repartitioned
energy given to particle $i$ in the interaction, chosen randomly
with a uniform distribution at every interaction; $E_{B}$ is the
energy of the particle from the bath with which particle $i$ may
interact, which at every interaction is chosen randomly from the
equilibrium distribution in the bath, $p_B(E_{B}) = T_{B}^{-1}
exp(-E_{B}/T_{B})$.

%%%%%%%%%%%%%%%%%%%%%%%%%%%%%%%%%%%%%%%%%%%%%%%%%%%%%%%%%%%%%%%%%%%%%%%%%
%Moments:
%%%%%%%%%%%%%%%%%%%%%%%%%%%%%%%%%%%%%%%%%%%%%%%%%%%%%%%%%%%%%%%%%%%%%%%%%

We first demonstrate how all moments of $p(E)$ may be evaluated
exactly from the dynamical rule for general restitution
coefficient $\alpha$ and coupling strength $f$. Later, we shall
consider the generating function of $p(E)$, using a procedure
which is formally possible for the general case, but which will be
fully solved only for limiting cases.

The first moment of $p(E)$ is the average energy, which in analogy
with granular materials is denoted as the granular temperature,
$T_G \equiv \langle E \rangle$. This is evaluated by averaging Eq.
(\ref{eq:dyn_rule}) over the whole system:
\bea T_G(t+dt) &=& (1- \Gamma dt)
T_G(t) + (1-f) \Gamma dt \alpha T_G(t) \non \\
&+& f \Gamma dt \half (T_G(t) + T_{B} ) \label{eq:1st_moment}.
\eea
In the steady state $T_G(t+dt)=T_G(t)$, and therefore
\be T_G = T_{B} / \left[2 \alpha -1 + 2 (1-\alpha)/f \right]
\label{eq:t_gran}. \ee
%
%\be T_G = \frac {T_{B}}{2 \alpha -1 + 2 (1-\alpha)/f}
%\label{eq:t_gran}. \ee
%

\begin{figure}[htb]
\includegraphics[width=8cm]{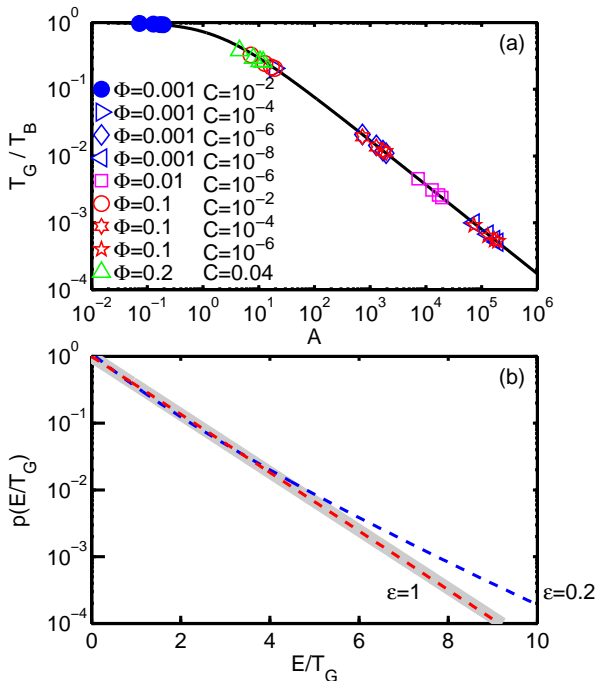}
\caption{\label{fig:compare_mf} a) The ratio $T_G/T_B$ vs. the
scaled parameter $A (\epsilon,\Phi,C)$, defined in the text.
Results from numerical simulations with different combinations of
the restitution coefficient $\epsilon$, the volume fraction $\Phi$
and the dimensionless coupling strength $C$ (symbols) agree with
the mean field calculation (solid line). Results with 400
particles are shown, however similar results were obtained with up
to 12800 particles. b) Normalized energy distributions from
simulations with $\Phi=0.1$ and $C=0.01$ (dashed lines), compared
to the Boltzmann distribution (thick gray line).}
\end{figure}

While the model described above is extremely simple, it is
interesting to note that it captures, at least qualitatively, some
aspects of an actual driven granular gas. In Eq. (\ref{eq:t_gran})
we see that $T_G$ is always smaller than $T_B$ and depends not
only on the dissipation through the restitution coefficient
$\alpha$, but also on the details of the coupling to the bath
through the coupling strength $f$ ($T_G$ coincides with $T_B$ only
for the two non-dissipative limiting cases: conservative
interactions ($\alpha=1$) and interactions only with the bath
($f=1$)). To see this for a real granular gas (in 2D, for
convenience), we estimate the ratio $T_G/T_B$  by the following
mean-field energy balance calculation (see also
\cite{cecconi_2004}). Consider a gas of grains of mass $m$,
diameter $D$, and restitution coefficient $\epsilon$, at volume
fraction $\Phi$. The mean time between collisions for a grain of
energy $E$ is $\frac{\pi D}{8\Phi} \sqrt{\frac{m}{2E}}$, and the
mean energy dissipated per collision is proportional to
$(1-\epsilon^2)T_G$. Consequently, the average energy loss rate
due to collisions is $\frac{K\Phi (1-\epsilon^2)}{D\sqrt{m}}
T_G^{3/2}$ where $K$ is a dimensionless constant. This is balanced
in the steady state by gain due to driving by, and loss due to
friction with the bath, $2 \gamma (T_B-T_G)$, yielding $T_B-T_G- A
T_B^{-1/2} T_G^{3/2} = 0$, which may be solved for $T_G/T_B$ as a
function of the single parameter $A (\epsilon,\Phi,C) \equiv
\frac{K \Phi }{2C}(1-\epsilon^2)$ (where we use the dimensionless
coupling strength $C \equiv \gamma D \sqrt{\frac{m}{T_B}}$). We
note that results of molecular dynamics simulations agree with
this estimate, as can be seen in Fig. \ref{fig:compare_mf}a. This
granular gas energy balance is more complicated than the
calculation leading to Eq. (\ref{eq:t_gran}), since the rate of
dissipative interactions in a granular gas depends not only on the
coupling strength $C$ and volume fraction $\Phi$, but is
dynamically determined by the typical energy $T_G$ through the
grain velocities. Moreover, density and energy correlations in
granular gases likely break the validity of such a mean field
approximation, especially at high volume fraction and low
restitution coefficient, or for higher moments of the energy.

For our model the frequency of dissipative interactions is
determined by the controllable parameter $f$, and since our model
is inherently correlation free, any moment of the energy may be
exactly calculated by taking the average of any power of Eq.
(\ref{eq:dyn_rule}). This yields the following recursion relation
for $\langle E^n \rangle$ in the steady state in terms of all
lower moments and the known moments of $E_B$, $\langle E_B^m
\rangle = m ! T_B^m $:
\bea \langle E^n \rangle = (n+1-f-2 \alpha ^n (1-f))^{-1} [ f
\langle
E_B^n \rangle \non \\ + \sum^{n-1}_{m=1} \left( \begin{array}{c} n \\
m \end{array} \right) \langle E^m \rangle \left(f \langle
E_B^{n-m} \rangle + (1-f)\alpha ^n \langle E^{n-m} \rangle
\right)] \label{eq:moments_recursion} .\eea
%

%%%%%%%%%%%%%%%%%%%%%%%%%%%%%%%%%%%%%%%%%%%%%%%%%%%%%%%%%%%%%%%%%%%%%%%%%
%Generating function:
%%%%%%%%%%%%%%%%%%%%%%%%%%%%%%%%%%%%%%%%%%%%%%%%%%%%%%%%%%%%%%%%%%%%%%%%%

In order to obtain a full solution of the energy distribution
$p(E)$, we introduce the generating function $g(\lambda) \equiv
\langle e^{-\lambda E} \rangle = \int_0^{\infty} e^{-\lambda E}
p(E) dE $. Averaging the exponential of Eq. (\ref{eq:dyn_rule})
over the whole system and considering the steady state yields
\bea \langle e^{-\lambda E} \rangle &=& (1- \Gamma
dt) \langle e^{-\lambda E} \rangle + (1-f) \Gamma dt
\int_0^1 \langle e^{- z \alpha \lambda E} \rangle ^2 dz \non \\
&+& f \Gamma dt \int_0^1 \langle e^{- z \lambda E} \rangle \langle
e^{- z \lambda E_B} \rangle dz \label{eq:gener_int_dynam}, \eea
or
\be g(\lambda)= (1-f)\int_0^1 g^2(z \alpha \lambda)dz + f \int_0^1
\frac{g(z \lambda)dz}{z \lambda T_B+1} \label{eq:gener_int}, \ee
where we have used the generating function of the exponential
energy distribution in the bath, $g_B(\lambda) \equiv
\int_0^{\infty} e^{-\lambda E_B} T_B^{-1} e^{-E_B/T_B} dE_B =
(\lambda T_B+1)^{-1}$. We now transform the integral equation
(\ref{eq:gener_int}) into a differential equation by change of
variables to $z \lambda$ in both integrals, multiplication by
$\lambda$ and differentiation by $\lambda$, yielding
\be \lambda \frac{dg(\lambda)}{d\lambda} = (1-f) g^2(\alpha
\lambda) + \left( \frac{f}{\lambda T_B+1}-1 \right) g(\lambda)
\label{eq:gener_diff}. \ee

\begin{figure}[htb]
\includegraphics[width=8cm]{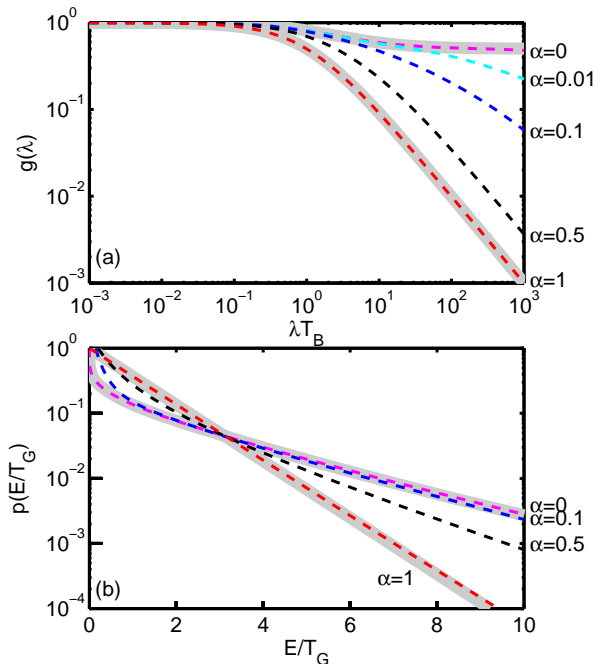}
\caption{\label{fig:gener_func} The generating function (a) and
normalized energy distribution (b) for coupling strength
$f=0.5$. Dashed lines are results of numerical simulations for
various values of the restitution coefficient $\alpha$, and thick
gray lines are theoretical solutions for the two limiting cases.}
\end{figure}

We first note that the generating function of the Boltzmann
distribution, $g_B(\lambda)$, solves Eq. (\ref{eq:gener_diff}) for
the two non-dissipative limiting cases $\alpha=1$ and $f=1$.
Furthermore, the solution for the limit of maximal dissipation
($\alpha=0$) is
$ g(\lambda)=  { _2F_1 }(1,2,2-f,- \lambda T_B) \cdot (\lambda
T_B+1) $,
%
%\be g(\lambda)=  { _2F_1 }(1,2,2-f,- \lambda T_B) \cdot (\lambda
%T_B+1) \label{eq:gener_a_0}, \ee
%
where $_2F_1$ is the Gauss hypergeometric function. The
theoretical results of these two limiting cases ($\alpha=0$ and
$\alpha=1$) are plotted in Fig. \ref{fig:gener_func} together with
results for $0 \leq \alpha \leq 1$ obtained from a Monte-Carlo
numerical simulation of the model with 1000 particles. The
numerical results indicate that the theoretical results known for
$\alpha=0$ and $\alpha=1$ bound the family of solutions for all
intermediate values $0<\alpha<1$. Note that the characteristic
form of the energy distribution in this exactly solvable model
qualitatively resembles that of granular gases (see Fig.
\ref{fig:compare_mf}b), for which only approximate solutions
exist.

%%%%%%%%%%%%%%%%%%%%%%%%%%%%%%%%%%%%%%%%%%%%%%%%%%%%%%%%%%%%%%%%%%%%%%%%%
%Fluctuation-dissipation relations:
%%%%%%%%%%%%%%%%%%%%%%%%%%%%%%%%%%%%%%%%%%%%%%%%%%%%%%%%%%%%%%%%%%%%%%%%%

We will now use FD relations in order to demonstrate one
observable deviation from equilibrium behaviour in dissipative
systems due to their non-Boltzmann energy distribution. In order
to measure FD relations in their simplest form we introduce an
internal degree of freedom per particle, $x_i$, which is coupled
to a uniform external field, $F$, so that the total energy of
every particle is $U_i=E_i-x_iF$, where $E_i$ is now referred to
as the kinetic energy. We consider the relation between the
fluctuation in $x$, $\langle \Delta x^2 \rangle \equiv \langle x^2
\rangle - \langle x \rangle ^2$, and its susceptibility with
respect to changes in $F$, $\chi \equiv \frac{\partial \langle x
\rangle}{\partial F}$.

We assume driven dissipative dynamics for the kinetic energy
together with non-dissipative exchange of kinetic energy and
internal energy, as described by the following equations of
motion,
\begin{subequations}\label{eq:dyn_rule_fd}
\bea E_i(t+dt)= \left\{
\begin{array}{cc}
  \underline{value}: & \underline{probability:} \\
  E_i(t) & 1- \Gamma dt \\
  z \alpha (E_i(t)+E_j(t))  & (1-f) \Gamma dt \\
  z (E_i(t)+E_{B})  & f (1-h) \Gamma dt \\
  z (E_i(t)-x_jF)  & f h \Gamma dt
\end{array} \right.\label{eq:dyn_rule_fd_e},
\eea
\bea x_i(t+dt)= \left\{
\begin{array}{cc}
  \underline{value}: & \underline{probability:} \\
  x_i(t) & 1- f h \Gamma dt \\
  z \left(x_i(t)-\frac{E_j}{F}\right)  & f h \Gamma dt
\end{array} \right.\label{eq:dyn_rule_fd_x},
\eea
\end{subequations}
where $h$ is a parameter introduced to describe the ratio between
interactions with the bath and interactions between the two types
of degrees of freedom. By calculating the first two moments of Eq.
(\ref{eq:dyn_rule_fd_x}) one sees that $\chi = \frac{\langle E
\rangle}{F^2}$ and $\langle \Delta x^2 \rangle = \frac{\langle E^2
\rangle}{2F^2}$. That is, the FD relation, $\langle \Delta x^2
\rangle  = \chi T_{FD}$, is satisfied with an effective
temperature $T_{FD}=\frac{\langle E^2 \rangle}{2\langle E
\rangle}$, which probes the distribution of the kinetic energy
$E$, determined from Eq. (\ref{eq:dyn_rule_fd_e}). Note that
$\chi$ and $\langle \Delta x^2 \rangle$ diverge as $F \rightarrow
0$ (where $T_{FD}$ is normally defined), however their ratio is
finite and independent of $F$.

$T_{FD}$, which characterizes fluctuations, results from the
second moment of the energy distribution, and is generally
different from $T_G$, which describes the first moment. $T_{FD}$
and $T_G$ coincide only if $\langle E^2 \rangle = 2 \langle E
\rangle ^2$, which is the case for the Boltzmann distribution. For
dissipative systems far from equilibrium $T_{FD}$ is larger than
$T_G$, and their ratio is given for our model in the limit
$h\rightarrow 0$ by
\bea \frac{T_{FD}}{T_G} = \frac{\langle E^2 \rangle}{2 \langle E
\rangle ^2} = [2(2-f)-2 \alpha (4-f)(1-f) \non \\ + \alpha ^2
(4-3f)(1-f)]/[f(3-f-2 \alpha^2 (1-f))] \label{eq:tfd_tg_ratio}.
\eea
In dissipative systems with strong coupling ($f \approx 1$) and
large restitution coefficient ($\alpha \approx 1$) the energy
distribution is close to exponential, hence the values of $T_{FD}$
and $T_G$ are similar (but not identical). Generally, the ratio
$T_{FD}/T_G$ may reach any value larger than one, and diverges for
vanishing coupling strength ($f \rightarrow 0$). As has recently
been predicted by kinetic theory \cite{garzo_2004}, we expect
$T_{FD}$ to be larger than $T_G$ in granular gases as well, where
the energy distribution is non-exponential. In the cases studied
numerically \cite{t_fd_gran} the energy distributions were only
slightly non-exponential, resulting in small differences between
$T_{FD}$ and $T_G$, which explains their seeming coincidence.

%%%%%%%%%%%%%%%%%%%%%%%%%%%%%%%%%%%%%%%%%%%%%%%%%%%%%%%%%%%%%%%%%%%%%%%%%
%Conclusions:
%%%%%%%%%%%%%%%%%%%%%%%%%%%%%%%%%%%%%%%%%%%%%%%%%%%%%%%%%%%%%%%%%%%%%%%%%

In conclusion, we have presented a simple dissipative model,
solved it in terms of all energy moments in the general case, and
obtained an exact expression for the generating function in the
maximally dissipative limit. Although our model is inspired by
granular gases, we believe it may have relevance to a broader
class of driven dissipative systems. A dissipative system coupled
to a heat bath is very different from a conservative system
coupled to the same bath. Not only is the granular temperature
lower than the bath temperature, but it also depends on the
coupling details. In addition, FD relations hold with an effective
temperature $T_{FD}$ which characterizes the second moment of the
energy distribution, and is larger than the granular temperature,
$T_G$. These exactly coincide in equilibrium, where the energy
distribution is exponential, however generally differ in
dissipative systems. We expect that careful numerical studies of
granular gases will show this heretofore unobserved difference.

%%%%%%%%%%%%%%%%%%%%%%%%%%%%%%%%%%%%%%%%%%%%%%%%%%%%%%%%%%%%%%%%%%%%%%%%%
%Acknowledgments:
%%%%%%%%%%%%%%%%%%%%%%%%%%%%%%%%%%%%%%%%%%%%%%%%%%%%%%%%%%%%%%%%%%%%%%%%%

We would like to thank Naama Brenner, Guy Bunin, Bernard Derrida,
J. Robert Dorfman, Jean-Pierre Eckmann, Dmitri Grinev, Fred
MacKintosh and Yael Roichman for helpful discussions. DL
acknowledges support from grant no. 88/02 of the Israel Science
Foundation and the Fund for the Promotion of Research at the
Technion.

%%%%%%%%%%%%%%%%%%%%%%%%%%%%%%%%%%%%%%%%%%%%%%%%%%%%%%%%%%%%%%%%%%%%%%%%%
%References:
%%%%%%%%%%%%%%%%%%%%%%%%%%%%%%%%%%%%%%%%%%%%%%%%%%%%%%%%%%%%%%%%%%%%%%%%%

%%%%%%%%%%%%%%%%%%%%%%%%%%%%%%%%%%%%%%%%%%%%%%%%%%%%%%%%%%%%%%%%%%%%%%%%%


\begin{thebibliography}{10}

\bibitem{granular_gases}
{\em Granular Gases}, edited by T. P\"oschel and S. Luding
(Springer, Berlin, 2001); {\em Granular Gas Dynamics}, edited by
T. P\"oschel and N. Brilliantov (Springer, Berlin, 2003).

\bibitem{incline_flow}
L.E. Silbert {\em et al.}, Phys. Rev. E {\bf 64}, 051302 (2001);
C. Ancey, Phys. Rev. E {\bf 65}, 011304 (2001); J. Rajchenbach,
Phys. Rev. Lett. {\bf 90}, 144302 (2003).

\bibitem{rotating_drum}
E. Cl$\acute{e}$ment, J. Rajchenbach and J. Duran, Europhys. Lett.
{\bf 30}, 7 (1995); G.H. Ristow, Europhys. Lett. {\bf 34}, 263
(1996); K. Yamane {\em et al.}, Phys. Fluids {\bf 10}, 1419
(1998). D. Levine, Chaos {\bf 9}, 573 (1999);

\bibitem{vibrated_container}
S. Luding {\em et al.}, Phys. Rev. E {\bf 49}, 1634 (1994); F.
Rouyer and N. Menon, Phys. Rev. Lett. {\bf 85}, 3676 (2000); D.L.
Blair and A. Kudrolli, Phys. Rev. E {\bf 64}, 050301(R) (2001);
D.L. Blair and A. Kudrolli, Phys. Rev. E {\bf 67}, 041301 (2003);

\bibitem{noije_ernst_1998}
T.P.C. van Noije and M.H. Ernst, Granular Matter {\bf 1},  57
(1998).

\bibitem{geisshirt_1998}
K. Geisshirt {\em et al.}, Phys. Rev. E {\bf 57}, 1929 (1998).

\bibitem{cecconi_2004}
F. Cecconi {\em et al.}, J. Chem. Phys. {\bf 120}, 35 (2004).

\bibitem{t_fd_other}
L.F. Cugliandolo, J. Kurchan and L. Peliti, Phys. Rev. E {\bf 55},
3898 (1997); A. Barrat {\em et al.}, Phys. Rev. Lett. {\bf 85},
5034 (2000); A. Barrat {\em et al.}, Phys. Rev. E {\bf 63}, 051301
(2001); H.A. Makse and J. Kurchan, Nature {\bf 415}, 614 (2002);
I.K. Ono {\em et al.}, Phys. Rev. Lett. {\bf 89}, 095703 (2002);
S.A. Langer and A.J. Liu, Europhys. Lett. {\bf 49}, 68 (2000); G.
D'Anna {\em et al}, Nature {\bf 424}, 909 (2003).

\bibitem{t_fd_gran}
A. Puglisi, A. Baldassarri and V. Loreto, Phys. Rev. E {\bf 66},
061305 (2002); A. Barrat, V. Loreto and A. Puglisi, Physica A {\bf
334}, 513 (2004).

\bibitem{maxwell_model}
E. Ben-Naim and P.L. Krapivsky, Phys. Rev. E {\bf 61}, R5 (2000);
E. Ben-Naim and P.L. Krapivsky, Phys. Rev. E {\bf 66}, 011309
(2002); D. ben-Avraham {\em et al.}, Phys. Rev. E {\bf 68},
050103(R) (2003); A. Santos and M.H. Ernst, Phys. Rev. E {\bf 68},
011305 (2003);

\bibitem{mackintosh_2004}
J.S. van Zon and F.C. MacKintosh, Phys. Rev. Lett. {\bf 93},
038001 (2004).

\bibitem{ulam_1980}
S. Ulam, Adv. Appl. Math. {\bf 1}, 7 (1980).

\bibitem{garzo_2004}
V. Garz$\acute{o}$, Physica A {\bf 343} 105 (2004).

\end{thebibliography}
\end{document}